\documentstyle[times,epsfig]{mn}
\input{epsf}

\newcommand{\bez}{\begin{eqnarray*}}
\newcommand{\eez}{\end{eqnarray*}}
\newcommand{\be}{\begin{equation}}
\newcommand{\ee}{\end{equation}}
\newcommand{\beq}{\begin{eqnarray}}
\newcommand{\eeq}{\end{eqnarray}}
\newcommand{\bc}{\begin{center}}
\newcommand{\ec}{\end{center}}

\newbox\grsign \setbox\grsign=\hbox{$>$} \newdimen\grdimen \grdimen=\ht\grsign
\newbox\simlessbox \newbox\simgreatbox \newbox\simpropbox
\setbox\simgreatbox=\hbox{\raise.5ex\hbox{$>$}\llap
     {\lower.5ex\hbox{$\sim$}}}\ht1=\grdimen\dp1=0pt
\setbox\simlessbox=\hbox{\raise.5ex\hbox{$<$}\llap
     {\lower.5ex\hbox{$\sim$}}}\ht2=\grdimen\dp2=0pt
\setbox\simpropbox=\hbox{\raise.5ex\hbox{$\propto$}\llap
     {\lower.5ex\hbox{$\sim$}}}\ht2=\grdimen\dp2=0pt

\def\simlt{\mathrel{\copy\simlessbox}}

\def\fuvx{F_{\mathrm{uv}}/F_{\mathrm{x}}}

\begin{document}

\title[spherical accretion flows]{A quasi-spherical inner accretion flow in Seyfert galaxies ?}

\author[J.~Malzac]{\parbox[]{6.8in} {Julien~Malzac}\\
Osservatorio Astronomico di Brera, via Brera, 28, 20121 Milan, Italy}

\date{Accepted, Received}

\maketitle


\begin{abstract}
 
 We study a phenomenological model for the continuum 
emission of Seyfert galaxies.  
In this quasi-spherical accretion scenario, 
the central X-ray source is constituted by 
a hot spherical plasma region surrounded by spherically distributed 
cold dense clouds.
The cold material is radiatively coupled with the hot thermal plasma. 
Assuming energy balance, we compute
 the hard X-ray spectral slope $\Gamma$ and reflection amplitude $R$.
 This simple model enables to reproduce  
both the range of observed hard X-ray spectral slopes,
and reflection amplitude $R$. It also predicts a correlation between 
$R$ and $\Gamma$ which is very close to what is observed. 
Most of the observed spectral variations from source to source,
 would be due to  differences in the cloud covering fraction.
If some internal dissipation process is active
in the cold clouds, darkening effects
 may provide a simple explanation for the observed distributions 
of reflection amplitudes, spectral slopes, and UV to X-ray flux ratios.

\end{abstract}


 
\section{Introduction}
\footnotetext{ E-mail: malzac@brera.mi.astro.it}

The hard X-ray spectra of galactic black holes (GBH) and radio-quiet active
 galactic nuclei (AGN) are generally thought to form through thermal
 Comptonisation process, i.e., multiple up--scattering of soft seed photons
in a hot plasma.
This process indeed, leads generally to a power law hard X-ray spectrum 
with a nearly  exponential cut-off at an energy characteristic 
of the plasma temperature.
The spectral slope depends both on the temperature $T$ and the
Thomson optical depth of the plasma $\tau$. The spectra are generally
 harder when both $\tau$ and $T$ are larger (e.g. Sunyaev \& Titarchuk 1980). 
Spectral fits indicate $kT\sim100$~keV and $\tau \sim 1$ (see reviews by Zdziarski et al. 1997, Poutanen 1998).

The nature of the comptonising  plasma is still unclear.
 This could be the hot inner part of the accretion disc powered by
 viscous dissipation (e.g. Shapiro, Lightman \& Eardley 1976).
 Another possibility would be a patchy corona lying 
atop an accretion disc and powered by magnetic reconnection 
(e.g.  Bisnovatyi-Kogan  \&  Blinnikov  1977; Haard,
 Maraschi \& Ghisellini 1994).
 See Beloborodov 1999b  (B99b) for a recent review. 
In both cases the main  dominant cooling  mechanism of 
the hot plasma is Compton cooling.
The soft cooling photons may be generated internally 
(e.g. by cyclo/synchrotron processes, see
 however Wardzinski \& Zdziarski 1999),
or more likely be produced by a cold medium in the vicinity 
of the hot plasma.  

In addition the spectra exhibit reflection features,  
 in  particular,  the Compton bump and the  fluorescent  iron
line which are the signatures for the presence of cold matter
which reflects the primary hard radiation (George \& Fabian 1991;
 Nandra \& Pound 1994). This cold matter could take
 the form of standard accretion disc 
or be constituted of cold clouds in the vicinity the hot source.

A correlation between the amplitude of the reflection component, $R$, and 
the photon index, $\Gamma$, is observed both in sample of sources and in the
 evolution 
of individual sources. This correlation is reported both in AGN
 and GBH (Zdziarski, Lubi\'nski \& Smith 1999, hereafter ZLS99; Gilfanov,
 Churazov \& Revnitsev 2000, hereafter GCR00).
The existence of a correlation
strongly support models
where the cooling of the plasma is due to thermal radiation from the same
medium that produces the reflection component 
(Haardt \& Maraschi 1993; ZLS99). It is an evidence 
against models where the soft photons are generated 
internally and independently of the reflection component.
It is also an evidence against models where most of the
reflection is produced very far from the hot source 
(e.g. on a torus;  Ghisellini, Haardt \& Matt 1994; Krolik, 
Madau \& \.Zycki 1994).

It could also be an argument against the patchy corona scenario
which, in its simplest version, produces an anti-correlation between 
$R$ and $\Gamma$ (e.g. Malzac, Beloborodov \& Poutanen 2001, hereafter MBP).
However, the energetics of the source may be strongly modified by
mildly relativistic bulk motions in the corona (Beloborodov 1999a).
The effects of bulk motion enable to reproduce both individual
 spectra and the observed $R$-$\Gamma$ correlation in AGN as well 
as in GBH (B99b; MBP).

The alternative model is a hot quasi-spherical plasma, surrounded by
 a cold accretion disc which can overlap with the
 hot inner region (e.g. Poutanen, Krolik \& Ryde 1997).  
 This latter model seems to fit well the $R$-$\Gamma$ correlation 
observed in galactic sources (GCR00, see however Beloborodov 2001).

In the case of Seyfert galaxies 
the measured $R$ values should be 
taken with caution since they may be affected by the 
larger uncertainties in the spectrum measurement.
It is however striking that the hot-sphere plus cold-disc  
model cannot account for the whole range of observed $R$
 in Seyfert galaxies (ZLS99).
 In particular, since the outer disc is always present, very low reflection
coefficients ($R<0.1$), observed in many sources, cannot be produced (unless
 the disc is strongly ionized).  
 Due to the low covering fraction of the
 cold material, large observed reflection amplitudes ($R>1$)
 cannot be reached.

A simple geometry for the cold material that may produce both large
 and very low covering fraction is a spherical geometry
 where cold clouds are distributed 
spherically, around the central hot plasma.
The possibility for the existence of cold clouds in the innermost 
part of the accretion flow was suggested by Guilbert \& Rees 1988.
It has been further studied in details and shown to be physically realisable 
(Celotti, Fabian \& Rees 1992; Kuncic, Blackman \& Rees 1996;
 Kuncic, Celotti \& Rees 1997).
The particular spherical geometry considered here was proposed 
 by Collin-Souffrin et al. (1996). Different variations on this model
 have been further studied,
in the context of multi wavelength 
variability and line formation. It has been shown to present a range of 
advantages regarding to the observations (e.g. Czerny \& Dumont
 1998;  Abrassart \& Czerny 2000, hereafter AC00; Collin et al. 2000).

In all these studies the authors focussed on the cold blobs
themselves and not on the problem of energy balance
 in the hot phase. The influence of the cold matter distribution
on the primary emission was not considered.
Indeed, a
 fraction of the cold clouds thermal radiation enters the central 
hot region. The heating of the hot plasma is thus balanced by 
the Compton cooling due to the incoming soft photon flux.
The resulting equilibrium temperature $T$, (and thus the 
emitted the spectrum) then depends mainly on the cloud distribution.

Here we will show that taking into account these
latter effects enable to reproduce 
the range of observed $R$ and $\Gamma$ as well as
 the correlation that is observed   
Seyfert galaxies. With some additional assumptions, 
the distribution of  $R$, and $\Gamma$
as well as the UV to X-ray flux ratios,
 may be explained by darkening effects.

In section~\ref{sec:estimates},  we present our assumptions and 
provide analytical estimates for
 $R$ and $\Gamma$ as predicted by the spherical accretion model.
Then, we will use these estimates to constrain the
 model with observations.
In section~\ref{sec:closeclouds}, we consider the case where the 
clouds thermal emission is due only to reprocessing of the hard
 X-ray radiation.
In section~\ref{sec:remoteclouds}, we study the effects of an eventual
additional  internal energy dissipation process in the clouds.
Finally, in section~\ref{sec:consideration}, 
we comment on several observational issues.

\begin{figure}
\centerline{\scalebox{.8}{\epsfig{file=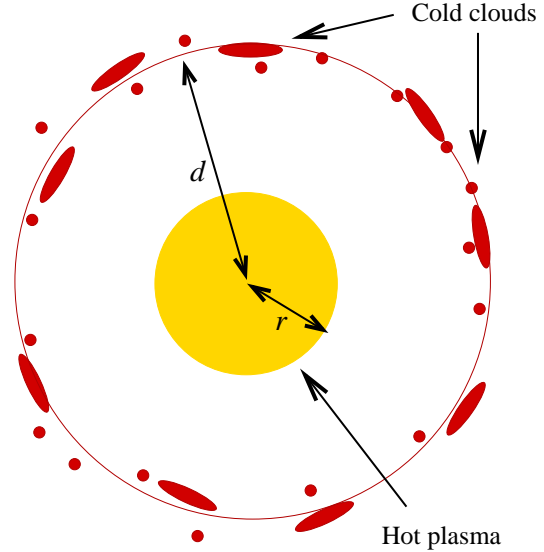}}}
\caption{The assumed geometry of the inner accretion flow: the black 
hole is at the centre of a spherical hot plasma cloud, with radius $r$,
 forming the hard X-ray source. Cold clouds  
are spherically distributed at some distance $d$ from the centre. They provide
the seed photons for Comptonisation in the hot phase.
 They intercept a fraction of the primary X-ray radiation. The main part of the intercepted X-ray flux is reprocessed as low energy (UV) radiation, the rest 
is reflected in the X-ray.
 The system is assumed to be in radiative equilibrium.  
\label{fig:geo}}
\end{figure}

\section{Quasi-spherical accretion model}\label{sec:estimates}

\subsection{Setting-up the picture}

 For the purpose of computing 
the spectral characteristics, the exact nature of the cold clouds 
does not need to be specified. They could consist in radially
 infalling material as well as the residual
 fragments from the inner part of a disrupted disc orbiting around 
the central object. We require the accretion flow to be
 spheric only in its inner parts.
 This does not preclude the existence of an outer accretion disc.
The clouds are even not required to be accreting
 and they could be part of an outflow or a wind.
They are however powered by accretion through reprocessing
of the primary hard X-ray radiation.

As long as the number of clouds is large enough
 the observed spectrum does not depend much on 
the particular cloud distribution
 or line of sight. Rather, it depends on general 
intrinsic characteristics of the source such 
as the cloud covering fraction and 
the average relative distance between 
the clouds and the hot comptonising plasma.
The shape of the individual clouds is then indifferent.

Let us consider that the hot phase constitutes a sphere with radius $r$.
We note $L_{\rm h}$, the power dissipated inside this hot sphere.
We assume that the cold clouds are all at the same distance $d$  
from the centre of the sphere (see Fig.~\ref{fig:geo}).
 We model their effects by assuming that 
 a fraction $C$ of the radiation impinging on the sphere
 of radius $d$ is intercepted.
 $C$ is thus the covering fraction of the clouds.
 A fraction $a$ of the intercepted flux is reflected while
 the rest is absorbed, reprocessed and reemitted as soft (UV) radiation.
$a$ is the energy and angle integrated hard X-ray albedo.  

 For simplicity, we will assume that all the \emph{reprocessed} power
is emitted by the \emph{inner} clouds surface,
i.e. \emph{toward the inner part of the accretion flow}.
This assumption is reasonable if the clouds are Compton thick 
and highly absorbing and thermal
 conductivity can be neglected. Then, most of the impinging energy 
is absorbed close to the inner surface and reemited locally.
In addition to reprocessed emission it is possible 
that the clouds radiate due to some additional internal
 dissipation process (see section~\ref{sec:remoteclouds}).
 We will allow for this possibility.
 We note $L_{\rm d}$ the total power dissipated in the cold clouds and 
radiated as UV radiation. Half of it is emitted outward, the other half
 is emitted toward the hot plasma.

 The virtual sphere of radius $d$ receives a total power
 $L_{\rm t}$. It arises from three contributions:
\begin{equation}
L_{\rm t}=L_{\rm x}+L_R+ L_{\rm uv},\label{eq:ltot}
\end{equation}
 respectively the comptonised luminosity, the reflected luminosity,
 the soft luminosity. A fraction $CL_{\rm t}$ is intercepted by the reflector 
and then reprocessed/reflected toward the inner direction.
An important quantity is the fraction of luminosity emitted 
inwardly by the reflector surface 
 which is not comptonised (i.e. emitted toward the reflector without 
interacting with the hot plasma):
\begin{equation}
K=1-\xi +\xi \exp{-\tau}.
\end{equation}
$\tau$ is the hot plasma Thomson optical depth,
 $\xi=\Delta\Omega/2\pi$ and  $\Delta\Omega$ is 
 the solid angle subtended by the hot
source as seen from the surface of the reflector.
In our spherical approximation, we have:
\begin{equation}
\xi=1-\sqrt{1-\left(\frac{r}{d}\right)^2}.
\end{equation}

In reality, one expects to have clouds 
radially distributed around the source, rather than at a given distance.
 Actually  quantities such as  
$d/r$, $\xi$ or $K$  should be regarded as \emph{average
 quantities}, indicative of the characteristic distance
 where most of the reprocessing takes place.
We expect that for most of possible radial distributions 
one can find an effective $d/r$ that provides a reasonable 
approximation to the high energy spectrum.
The only important effect of a very extended radial
 distribution would be that a significant fraction of
 the radiation reprocessed by the outer clouds, rather 
than feeding the hot phase, escapes trough reflection
 on the dark side of the innermost clouds. Taking into account
 this effect would require to introduce a transmission 
factor for reprocessed radiation which would depend on 
the specific cloud distribution. In order to limit the number 
of model parameters and for the sake of simplicity,
 we will not consider this possibility.

\subsection{Model parameters}

In the next sections  we will link the intrinsic physical
 and geometric properties of the source to observable quantities. 
Similar estimates for the luminosity ratios
and the amplification factor
 can be found in a different form in AC00.
We extend their results to the case where some dissipation occurs
 in the cold phase.
But the main improvement brought by the present work
 is to provide estimates for the two measured
 quantities that are commonly used to describe the hard X-ray spectra, 
namely the spectral slope $\Gamma$ and the reflection coefficient $R$.

These spectral characteristics are fully determined by 6 parameters:
 the covering factor $C$, the relative cloud distance $d/r$,
 the relative dissipation in the clouds $L_{\rm d}/L_{\rm h}$,
 the hot plasma scattering optical depth $\tau$ and the cloud albedo
 $a$ and the characteristic energy of the soft seed photons, which 
may be quantified using an effective blackbody temperature $T_{\rm bb}$.

The albedo depends mainly on the ionization state of matter
 and also on the shape of the incident spectrum.   
In the following, we will assume that the cold matter in the clouds 
is neutral.  The characteristic values of the albedo for typical
 Comptonisation spectra in AGN is then $a\sim0.1$ (see MBP).
Our results are insensitive to its actual value
 as long as it is low ($a<0.3)$. 
Since this condition is always realized for neutral matter we will fix
 $a$$=$0.1 without considering any dependence 
on the spectral parameters. 
On the other hand, there is a complication due to our particular
 geometry which make possible multiple reflections inside
 the spherical cavity.
The neutral albedo for a reflection-like spectrum differs significantly 
from that for the primary Comptonisation spectrum. We will thus consider
 a different albedo, noted  $a_{R}$, for the reflected luminosity. 
 A Monte-Carlo simulation, considering a typical reflection
 spectrum incident on neutral matter, provided the value $a_{R}=0.4$ used below.

Precise estimates of $\tau$ from spectral analysis are
 quite model dependent.
 In the same object one can infer $\tau$ being as low as a few tenth, 
and up to a few, depending on the fitting model (Petrucci et al. 2000)
 and may differ from source to source.
However the 2-10 keV spectral slope $\Gamma$ depends only weakly on $\tau$ 
(see section~\ref{sec:spectralslo} and Fig.~\ref{fig:ampli}).
 The main effect of $\tau$ is on the reflection amplitude $R$
 due to the destruction of the reflection
 component crossing the hot phase (MBP).
This effect is important only when the clouds 
 are extremely close to the source (i.e. $d/r < 1.1$; see
 section~\ref{sec:rdec} and Fig.~\ref{fig:rgdec}).
Since, $\tau$ is generally found to be of order unity, we will fix 
 the optical depth to the standard value $\tau$=1.

The value of the effective blackbody temperature, 
$T_{\rm bb}$ affects $\Gamma$ (and also slightly $a$).
Changes in $kT_{\rm bb}$ have however almost no effects,
 as long as they are kept inside 
the observed range where the bulk of the soft luminosity emerges in AGN
 (5-50 eV; see Fig.~\ref{fig:ampli}).
We will thus assume that $T_{\rm bb}$  lies
 somewhere in that range.

The number of effective parameters is thus reduced to 3,
 namely: $d/r$, $C$ and $L_{d}/L_{h}$.

\subsection{The reflected component}

The reflected luminosity $L_R$ impinging on the sphere of radius $d$
 has been emitted by the reflector itself from a fraction of the impinging 
X ray luminosity and crossed the system without being
 Compton scattered, we can thus write:
\begin{equation}
L_R=KC(aL_{\rm x}+a_{R}L_R). \label{eq:LR}
\end{equation}
The additional $L_{R}$ term on the right hand side of equation~\ref{eq:LR}
 accounts for multiple reflections in the cloud system.

The amplitude of reflection $R$ is usually defined by the ratio of an 
observed reflection component to that expected from an isotropic 
point source with the same luminosity illuminating an infinite slab:
\begin{equation}
R=\frac{L_{R}}{L_{R}^{slab}}.
\end{equation}
\noindent In our case the latter can be estimated as:
\begin{equation}
L_{R}^{slab}\sim a \frac{L_{\rm x}}{2} f(\cos i). \label{eq:refiso}
\end{equation}

Indeed, in the disc illumination framework, half of the hard X-ray
 luminosity is emitted toward the disc, of which a fraction $a$ is then
 reflected. The angle $i$ is the assumed inclination of the
 slab. The factor $f(\cos i)$ represents the angular
 dependence of reflection in the reference slab model. 
 We estimate the $f$ function using the approximation
 given by Ghisellini et al. (1994):
\begin{eqnarray}
f(\mu)=\frac{3\mu}{4}[(3-2\mu^{2}+3\mu^{4})\ln\left(1+\frac{1}{\mu}\right)+\nonumber\\
(3\mu^{2}-1)\left(\frac{1}{2}-\mu\right)].
\end{eqnarray}

It is then straightforward to relate the observable $R$ 
to the physical parameters of the spherical model:
\begin{equation}
R\sim\frac{L_R}{L_{\rm x}}\frac{2}{a f(\cos i)}=\frac{2CK}{1-a_{R}CK} 
f^{-1}(\cos i).\label{eq:spherR}
\end{equation}

Usually $i$ is a frozen parameter of the fitting model
enabling to measure $R$ from comparison to the slab reflection model
(cf {\sc pexrav} model in {\sc XSPEC}, Magdziarz \& Zdziarski 1995).
In sections~\ref{sec:closeclouds} and \ref{sec:remoteclouds}, we will compare the model predictions with the 
the observational $R$ values given by ZLS99. These data have been
 fitted with $i$$=$$30^{\circ}$. 
We will thus adopt this value in most of our numerical estimates.

\subsection{The amplification factor}

The soft flux incident on the virtual sphere of radius $d$, 
arises partly from reprocessing on the reflector itself and partly
 from internal dissipation:

\begin{equation}
L_{\rm uv}=KC\left[(1-a)L_{\rm x}+(1-a_{R})L_R+L_{\rm uv}\right]+KL_{\rm d}/2.\label{eq:luv1}
\end{equation}
\noindent The global energy balance can be written:
\begin{equation}
L_{\rm h}+L_{\rm d}/2=(1-C)L_{t},\label{eq:globalenergybalance}
\end{equation}
where $L_{\rm h}$ and $L_{\rm d}$ are the power dissipated in the hot and cold phase respectively.
Combining equations~\ref{eq:ltot}, \ref{eq:luv1} and \ref{eq:globalenergybalance} and leads to:
\begin{equation}
L_{\rm uv}=\frac{KC (\omega-a)L_{\rm h}+K(\omega-aC)L_{\rm d}/2}{(1-C)(1-CKa_{R})}, \label{eq:luv}
\end{equation}
where we set~: 
\begin{equation}
\omega=1-KC(a_{R}-a).
\end{equation}

We defined $L_{\rm uv}$ as the soft luminosity impinging on 
the virtual sphere. $L_{\rm uv}$ can also be considered as the amount of
 soft luminosity, inwardly emitted by the reflector, that reaches
 the sphere of radius $d$ without interacting in the hot plasma.
 The total soft luminosity emitted by the reflector is thus $L_{\rm uv}/K$.  
The soft luminosity entering the hot phase is then:
\begin{eqnarray}
L_{\rm s}=\frac{\xi}{K}L_{\rm uv}.\label{eq:ls}
\end{eqnarray}
The total luminosity escaping from the hot phase is the sum of the power
 dissipated internally and the total power received from the cold phase:
\begin{equation}
L=L_{\rm h}+C\xi L_{t}+\xi L_{\rm d}/2.
\end{equation}
Then using equation~\ref{eq:globalenergybalance}, it is straightforward to show that
\begin{equation}
L=\frac{(1-C+C\xi)L_{\rm h}+\xi L_{\rm d}/2}{1-C}.
\label{eq:lhot}
\end{equation}
Finally, combining equation~\ref{eq:lhot} with \ref{eq:luv} and \ref{eq:ls},
we derive an estimate for the amplification factor:
\begin{equation}
A=L/L_{\rm s}=\frac{(1-KCa_{R})\left[(1-C+C\xi)L_{\rm h} + \xi L_{\rm d}/2\right]}{\xi C(\omega-a)L_{h}+\xi(\omega-aC)L_{\rm d}/2}. \label{eq:spherampl}
\end{equation}

\subsection{The spectral slope}\label{sec:spectralslo}
\begin{figure}
\centerline{\scalebox{1}{\epsfig{file=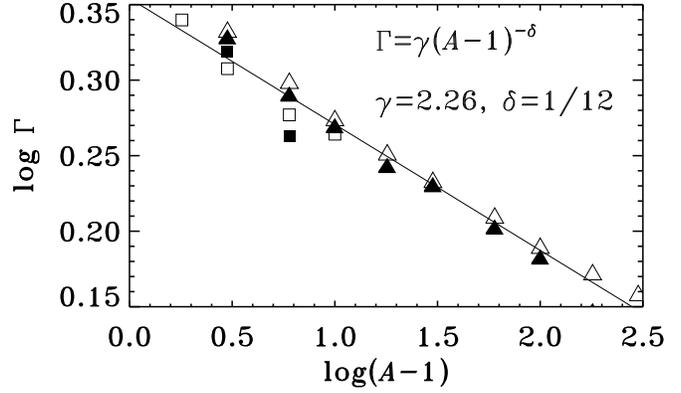,width=8.5cm,height=5cm}}}
\caption{Relation between amplification factor and photon index $\Gamma$,
for a spherical geometry. Soft photons are injected at the sphere surface, 
with a blackbody spectrum of temperature $T_{\rm bb}$.
The symbols show the results of non-linear Monte-Carlo simulations.
 Open and filled symbols respectively stand for $T_{\rm bb}$=5 and 50  eV.
 The radial Thomson optical depth of the sphere was fixed to
 $\tau$=1 (triangles), and $\tau$=0.2 (squares).
The solid line shows
the approximation given by
 equation~\ref{eq:total} with $\gamma$=2.26 and $\delta$=1/12.
\label{fig:ampli}}
\end{figure}  
A commonly measured quantity is the photon index $\Gamma$. 
It represents the spectral slope of the \emph{intrinsic} Comptonisation
 power law spectrum (i.e. not including reflection). It is generally measured
in the energy range 2--10 keV for a flux given in units of
photon per energy (e.g. photon keV$^{-1}$~cm$^{-2}$~s$^{-1}$).      

The amplification factor is directly related to the photon index.
The $\Gamma$ versus $A$ relation may be derived from numerical simulations.
We use the Monte-Carlo code of Malzac \& Jourdain (2000) 
based on the non-linear method proposed by Stern et al.~(1995).
We consider a sphere of thermal plasma
 with optical depth $\tau$ defined along its radius.
The sphere is homogeneously heated with a power $L_{\rm h}$. 
Soft photons are \emph{externally} injected at the sphere surface
 with a black body spectrum of temperature $T_{\rm bb}$.
 The injected soft luminosity is $L_{\rm s}$.
The density is homogeneous inside the sphere.
It is however divided in 10 radial zones in order to account for
 temperature gradients.
For a fixed ratio $L_{\rm h}/L_{\rm s}$=$A$-$1$, the code computes the equilibrium
 temperature structure as well as the spectrum of escaping radiation.
We then use a least-square fit to the 2-10 keV spectrum 
to determine the photon index $\Gamma$. 

Simulations were performed for several values of $A$-$1$ in the range 1--1000,
 for 
 $T_{\rm bb}$$=$ 5 and 50 eV, $\tau$$=$0.2 and 1. We then discarded simulations
resulting in a volume averaged electron temperature larger than 400 keV,
i.e. larger than what is inferred from the observations of the high energy cut-off.
We also discarded simulations resulting in $\Gamma$ outside of the range of interest 
regarding to the observations (1.4--2.2).
 As shown in Fig.~\ref{fig:ampli}, this sample of numerical
 results can be represented 
by the functional shape proposed by B99b:
\begin{equation}
\Gamma= \gamma(A-1)^{-\delta}, \label{eq:total}
\end{equation}
with  $\gamma$=$2.26$ and $\delta$=$1/12$.

The largest error is about 7 per cent, 
 arising for $T_{\rm bb}$$=$50 eV, $L_{\rm h}/L_{\rm s}$$=$6 and $\tau$$=$0.2  ($T$$=$360 keV).
If one except this high temperature case the error is less than 5 per cent.
We conclude that, at least for $\tau$ in the range 0.2--1, and all the other parameters
being in the observed range, $\Gamma$ depends (almost) only on the amplification factor.
This dependence can be represented with a good accuracy by equation~\ref{eq:total}.

The value of the coefficient differs slightly from those
originally given by B99b ($\gamma$$=$2.33, $\delta$$=$$1/10$).
The reasons for these small differences  
 could be that we performed a detailed	
 radiative transfer for the sphere geometry,
 while Beloborodov's results are based on 
a simple escape probability formalism 
in a one zone approximation (Coppi 1992),
 and photons injected at the centre.
Another possible cause for the discrepancy, is that we
fitted numerical results allowing temperature to adjust
while temperature was fixed in the calculations of B99b and $\tau$ was varied according 
to the amplification.

Relation~\ref{eq:total}, used together with 
equation~\ref{eq:spherampl} 
relates the observable $\Gamma$ to the model parameters.

\begin{figure*}
\centerline{\epsfig{file=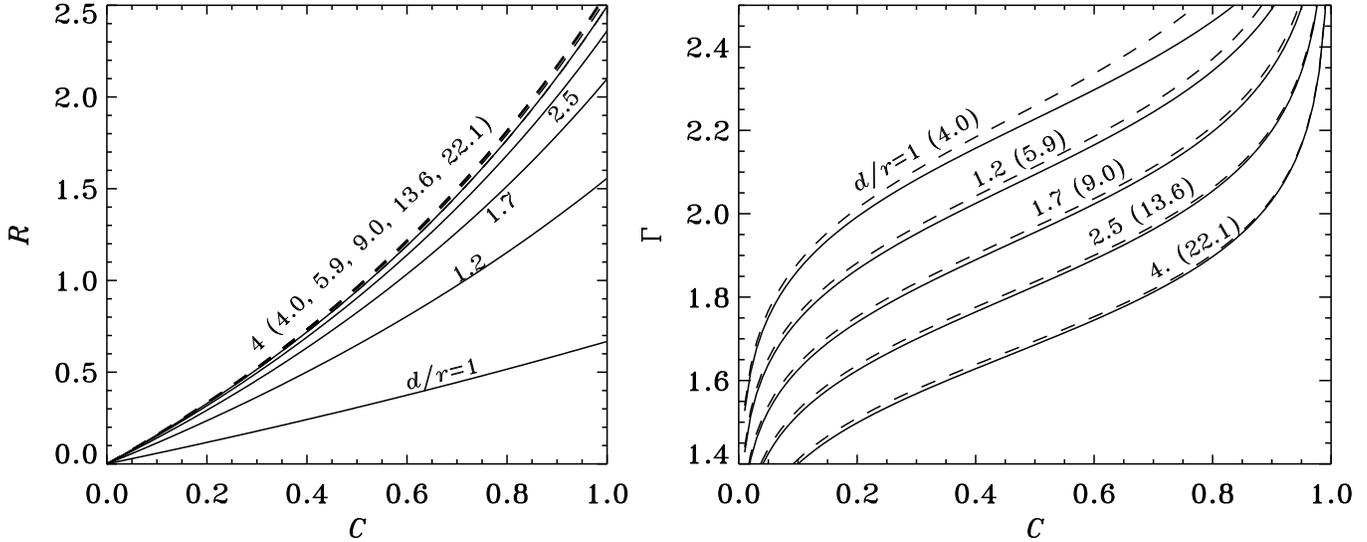,width=17.5cm,height=7cm}}
\caption{Dependence of the spectral parameters on the covering factor $C$.
Left panel: $R$ vs $C$; right panel: $\Gamma$ vs $C$.
The solid lines stand for the model without dissipation in
 the cold clouds ($L_{\rm d}$=0), and different distances:
$d/r=1, 1.2, 1.7, 2.5$ and $4$ respectively from top to bottom in right
 panel, from  bottom to top in left panel.
Dashed lines are for the model with intrinsic dissipation in the clouds
 such that $L_{\rm d}/L_{\rm h}=60C$ (see section~\ref{sec:remoteclouds}).
 Distances are $d/r= 4.0, 5.9, 9.0, 13.6, 22.1$ from top to bottom
 in right panel. In left panel all dashed curves are almost 
indistinguishable since $R$ is almost independent on $d$ for $d>2$.
In all curves the Thomson optical depth is $\tau$=1, 
cloud albedo is $a$=0.1,
 reflection angle parameter is $i$=$30^{\circ}$ (see section~\ref{sec:estimates}).\label{fig:rgdec}}
\end{figure*}

\subsection{The UV to X luminosity ratio}

The observed UV to X-ray flux ratio can be estimated as follow:
\begin{equation}
\frac{F_{\mathrm{uv}}} {F_{\mathrm{x}}}=\frac{(1-C)L_{\rm uv}+L_{\rm d}/2}{(1-C)(L_{\rm x}+L_R)}\label{eq:luvtolx}.
\end{equation}
Then, using equations~\ref{eq:luv1} and \ref{eq:luv},
 one can link the luminosity ratio to the model parameters:

\begin{equation}
\frac{F_{\mathrm{uv}}}{F_{\mathrm{x}}}=
\frac{
\left(\omega-a\right)KCL_{\rm h} +
\left[1+K\omega-KC(a_{R}+a)\right]L_{\rm d}/2}
{\omega\left[(1-KC)L_{\rm h} +
(1-K)L_{\rm d}/2\right]}
\label{eq:fuvsfx}.
\end{equation}
When comparing to observations one should keep in mind that this 
relation gives actually only a lower limit on the observed 
UV to X-ray flux. Indeed, there may be other sources of UV 
radiation in real objects (like an external accretion disc).


\section{Clouds without internal dissipation}\label{sec:closeclouds}

In this section, we will assume no dissipation in the blobs
 ($L_{\rm d}$=0).
The solid lines in Fig.~\ref{fig:rgdec} show $R$ (left panel)
and $\Gamma$ (right panel) as function of $C$ for different ratios $d/r$.

\subsection{$R$ versus $C$ relation }\label{sec:rdec}

The reflection amplitude grows almost linearly with $C$,
 starting from $R$$=$0 where there is no reflector ($C$=0), and
 reaching values of order $2$ for fully obscured source ($C\sim1$).
 Indeed, as seen from the source, the reflector then 
covers a solid angle $\sim 4\pi$. 

As the observed reflection component is produced on the inner side 
of the clouds, a fraction of the reflected photons
may be Compton scattered in the hot plasma before escaping to the observer.
The net effect is to decrease the apparent amount of reflection 
in the observed spectrum (see MBP). This Compton smearing effect
is obviously stronger at low $d/r$, when the hot source subtends a
larger solid angle. The maximum $R$ value is then lower, as shown on
left panel of Fig.~\ref{fig:rgdec}.
 This maximum value depends on the plasma 
optical depth $\tau$. For an optically thin plasma
 there is no significant reduction of
 $R$, while for very thick plasma and very close clouds ($d/r<1.1$) 
the reflection component can be almost totally destroyed. 

 On the other hand at large $d/r$ ($>2$), this effect can be 
neglected.
For large covering factors, multiple reflections on the inner 
side of the clouds tend to increase $R$ up to values larger than 2.
In this limit of $d/r$ larger than a few,
 the shape of the relation $R$ vs $C$ becomes fully independent of $d/r$.

In any case, for reasonable optical depths,
 the range of achievable $R$ when $C$ varies from 0 to 1,
is comparable to the range of observed $R$ (from 0 up to a few).

\subsection{$\Gamma$ versus $C$ relation} \label{sec:gdec}

For very low covering factors ($C\sim 0$) the amplification factor 
is huge and the spectrum is thus very hard.
 It softens very quickly with increasing $C$. At $C\sim0.1$ the growth rate 
decreases and $\Gamma$ then evolves slowly and almost linearly.
The spectral slope softens by $\Delta\Gamma\sim 0.6$
 between $C$=0.1 and $C$=0.8.
Then, for higher covering factors, the soft radiation emitted by the clouds
remain trapped inside the reflector sphere.
 This effect is non linear, a small increase in $C$ then increases 
dramatically the amount of soft cooling radiation $L_{\rm s}$ entering the hot plasma. As a consequence the emitted spectrum softens very quickly and $\Gamma$
diverges again for $C\sim1$.

The effect of changing $d/r$ does not change significantly 
the global shape described above. Increasing $d/r$ simply
 shift the curve downward. Indeed, the feedback from
 the clouds decreases, as they are more distant,
 and the spectra are harder on average.

For clouds distances in the range $r$$<$$d$$<$$4r$ the plateau of the curve
falls in the range $1.4<\Gamma<2.2$. The observed range of
$\Gamma$ values is then produced with intermediate $C$ values ($0.1<C<0.8$).

\subsection{$R$ versus $\Gamma$ relation\label{sec:rgammarel}}

As both $R$ and $\Gamma$ increase with $C$,
 it is obvious that changes in the covering factor
 will produce a correlation between these two spectral parameters.
It is then interesting to compare this correlation with the observed one.

The solid lines in Fig.~\ref{fig:spher_ldproptoC} show the $R$-$\Gamma$
correlation obtained when varying the covering factor of the source.
 The data points (from ZLS99) are also plotted in this figure.
The different curves are for different fixed values of the distance 
$d$ from the centre.
 Clearly, the overall shape of the correlation is well reproduced
 as long as the reflector is closer than 3 times the size of 
the X-ray source. Most of the data points lie in this range.
This range of distances could be the dispersion of
 the average distance of the reprocessor in different sources.
 It is also probable that most of the apparent spread is
 simply due to errors in the measurements of $R$ and $\Gamma$. 

If the reflector is located  at distance larger than $\sim 3r$ 
the feedback from reprocessing becomes too low and this produces
 very hard spectra. 
The distance is thus strongly constrained to be below $3r$.
If some dissipation occurs, however, a wider range of distances is allowed, 
see section~\ref{sec:remoteclouds}.

The prediction of a correlation, as well as its general shape,
 are insensitive to our model assumptions.

\subsection{Distributions in the $d/r$-$C$ plane}\label{sec:rcplane}

The system formed by equation~\ref{eq:spherR}, \ref{eq:spherampl} and 
\ref{eq:total}, giving  $R$ and $\Gamma$
as a function of $d/r$ and $C$ is, in general, invertible analytically.
 It is thus possible to transpose the 
$R$-$\Gamma$ data shown in Fig.~\ref{fig:spher_ldproptoC} into the $d/r$-$C$ plane. 
The result is shown in Fig.~\ref{fig:dsrc}. It illustrates the fact that the values 
 of the distance are comparable from source to source,
 while the covering factor can change by a large amount.
In the passive case the average distance is $<$$d/r$$>$=1.7, 
the root mean square spread is about 35 per cent.

For a fixed $L_{\rm d}/L_{\rm h}$, the 
correspondence between any couple ($R$, $\Gamma$ ) and ($d/r$, $C$)
 is univocal as long as $R$$\ne$0.
If $R$$=$0 then $C$$=$0, and $d/r$ is undetermined. 
 Several data points of our sample fall into this unphysical case.
The probable reason is that the reflection component was too weak
 to be detected.
Then we can infer that $C$ is actually very low, but 
$d/r$ cannot be estimated at all.
Therefore such data are not represented in Fig.~\ref{fig:dsrc}.

On the other hand, we used them for the purpose of computing the observed
 distribution of covering factors $p(C)$, displayed in Fig.~\ref{fig:pdec}. 
Clearly, the $C$-distribution appears to be a decreasing function of $C$. 
As shown in Fig.~\ref{fig:pdec}, it is reasonably well represented by 
a linear distribution i.e.:
\begin{equation}
p(C)=2(1-C). \label{eq:pdec1}
\end{equation}
This distribution has the same average value as the observed one:
 $<$C$>$=0.33.
The distribution $p(C)$ represents the probability density 
of observing a source with a covering factor $C$.
It does not necessarily reflects
 the \emph{intrinsic} distribution of covering factors,
 since $p(C)$, the distribution of \emph{observed} $C$,
 is likely to include some selection effects.

\subsection{$R$ and $\Gamma$ distributions} \label{sec:rgdis}

Assuming a fixed  $d/r$, for a given  $C$-distribution $p(C)$
 the respective distribution of $R$ and $\Gamma$ are respectively given by~:
\begin{eqnarray} 
p_{R}=p(C)\left(\frac{{\rm d}R}{{\rm d}C}\right)^{-1},\nonumber\\
p_{\Gamma}=p(C)\left(\frac{{\rm d}\Gamma}{{\rm d} C}\right)^{-1}.\label{eq:reflection}
\end{eqnarray}
For a given $p(C)$, an analytical expression for $p_{R}$ 
and $p_{\Gamma}$ can be easily derived 
from equations~\ref{eq:spherR} and~\ref{eq:spherampl}.

The solid lines in Fig.~\ref{fig:distrib},
 compare the predicted $p_{R}(R)$ and $p_{\Gamma}(\Gamma)$ for $p(C)$
 given by equation~\ref{eq:pdec1} with the observed distribution.
They are qualitatively reproduced for $d/r$$=$$<$$d/r$$>$=1.7.
This confirm that the actual  $p(C)$ is well represented by the 
equation~\ref{eq:pdec1}.

It is striking to have such a good
 agreement despite we neglected the spread in $d/r$.
Actually, the shapes of $p_{R}(R)$ and $p_{\Gamma}(\Gamma)$ 
are not very sensitive to the value of $d$, as long as it is kept
 in the observed range. A reasonable agreement is obtained as long as 
1.3$<$$d/r$$<$2.
 Only for very close reflector, the effects of Compton smearing 
of the reflection component become important. The $R$ distribution 
steepens slightly, with, on average  lower $R$ values.
Concerning $p_{\Gamma}$, at  larger $d$, the average $\Gamma$
 is harder, and the distribution is slightly narrower. 

Unlike the  $C$-distribution, the
theoretical $\Gamma$ distribution depends somewhat on the exact
 relation between $\Gamma$ and the amplification factor $A$.
 Different approximations used to represent this exact relation
may give different results.
 For example, if instead of equation~\ref{eq:total}
 with $\gamma$$=$2.26,  $\delta$$=$1/12,
 we use the coefficients given by B99b
(i.e. $\gamma$$=$2.33,  $\delta$$=$1/10)
we get a larger width and softer spectra.
 If instead we use $\gamma$$=$2.15,  $\delta$$=$1/14,
 as given by MBP for a cylindric geometry, we get,
 on the contrary, a distribution with a smaller width,
 peaking at harder spectra. 
In both cases, the shape of the distribution 
is still in qualitative agreement with the data.

We conclude that the spread in $d/r$ is actually negligible
and the $C$-distribution plays a major role for
 the observed $R$ and $\Gamma$ distributions.
Simply by setting the only significant parameters (namely $d/r$)
 to a value relatively close to 1.7, the model qualitatively reproduces
three observational pieces of information that
we have on the sources: the correlation $R$-$\Gamma$,
and the respective distributions of $R$ and $\Gamma$.

\subsection{Distribution of UV to X flux ratios $\fuvx$\label{sec:fuvx}}

Using equation~\ref{eq:fuvsfx}, one can derive the distribution of the 
$\fuvx$ that is predicted by the model with the $C$-distribution given by
 equation~\ref{eq:pdec1}.
The solid line in Fig.~\ref{fig:luvtolx}
shows the theoretical distribution which
is compared with the observed one (data from Walter \& Fink 1993).
This kind of comparisons should be taken with caution, since the expression for $\fuvx$ given by 
equation~\ref{eq:fuvsfx} refers to \emph{integrated quantities} 
while the measurements of Walter \& Fink were performed at a fixed UV and X-ray wavelengths 
(respectively 1375 \AA \/ and 2 keV). Moreover, as shown by Kuncic et al. (1997), small clouds 
probably emit in the EUV, rather than in the UV.

Despite these uncertainties,
 Fig.~\ref{fig:luvtolx} clearly shows that the modeled
 and observed distributions do not match.
For $\fuvx$ larger than a few, 
the model probability density is very close to 0. 
In contrast, the observed distribution extends up to quite large 
large UV to X flux ratios ($\fuvx\sim 50$).
Clearly, the inner clouds do not emit enough UV radiation to reach the highest
observed UV to X luminosity ratios. 
Note that, in principle, the model can produce arbitrarily large $\fuvx$  (see equation~\ref{eq:fuvsfx}),
 by setting $C$ and $K$ close to unity.
The problem is that it produces a too small amount of large $\fuvx$ sources.
Fitting the observed $\fuvx$ distribution would require 
a distribution of covering factor which would be a growing function of $C$, 
in contradiction to what is inferred from the X-ray data.

In an attempt to solve the problem, one can think about considering
 that a fraction of the reprocessed flux is 
 transmitted through the cloud instead of being fully emitted toward the hot plasma
 (see AC00). However, this does not increase the observed 
$\fuvx$ ratio. In fact, the opposite occurs. 
Due to the additional losses, amplification of the UV field in the cloud
 cavity is less efficient and, for a given covering factor,
the observed $\fuvx$ ratio is lower.

This apparent discrepancy can be overcomed if one consider 
that we assumed that all the soft radiation
 is produced trough reprocessing of 
the hard X-rays, \emph{without any additional source of UV radiation}.
The data simply suggest that such additional sources of UV radiation are 
present in real sources. These sources of UV radiation could be external,
 and not directly related to the central cloud system. 
For example, an external disc could be responsible
 for a significant (if not the largest) fraction of the observed 
UV emission. Then, the expected distribution UV to X luminosity
 ratio will depend on additional parameters such as the disc
 inner radius or inclination. Estimating the
$\fuvx$ distribution would require
 additional assumptions and several physical complications. 
From a qualitative point of view however, it is
 obvious that the resulting distribution 
would extend up to larger $\fuvx$, 
as required by the data.
 
Such an outer source of UV radiation is unlikely to contribute
sensibly to the cooling of the central source neither
to the reflected component. Thus, our estimates
 for $R$, $\Gamma$, $p_{R}$ 
and $p_{\Gamma}$ are still valid, as well as our conclusion drawn 
from the comparison with data: the present model enable to explain the main
 properties of the hard X-ray continuum observed in Seyfert galaxies
 provided that the passive clouds are distributed 
at distances $d$, in the range $r$--3$r$ from the central hot plasma region
 of radius~$r$.

Another possibility is that the additional source of UV luminosity
 would be located in the cold clouds themselves.
 Then, the excess soft luminosity 
may affect the radiative equilibrium and subsequently the emitted spectrum.
We thus have to take it into account when computing the spectral 
characteristics and distributions. This, in turn,
 requires some additional assumptions on the energy dissipation 
process presumed to be active in the clouds.
Then however, it makes it possible to build a simple self-consistent
 model accounting 
for the UV to hard X-ray spectral properties of the continuum emitted
 in Seyfert galaxies. This is the aim of the next section.

\begin{figure}
\centerline{\epsfig{file=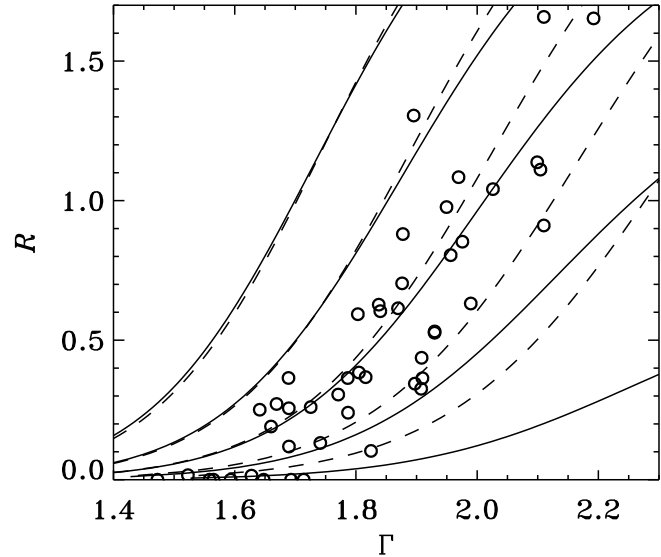,width=8.5cm,height=7cm}}
\vspace*{0.3cm}
\caption{The predicted $R$-$\Gamma$ correlation compared 
with observations of Seyfert galaxies.
The circles are the best fits parameters to the {\it Ginga} data
 (Zdziarski et al. 1999) in the $R$
 vs $\Gamma$ plane. The lines represent the correlation obtained for cold
 blobs are at a fixed distance from the hot plasma and covering fraction 
varying from 0 to 1. The solid lines stand for models with negligible dissipation in the
 blobs, from left to right for $d/r$=4, 2.5, 1.7, 1.2, and 1, respectively.
The dashed lines stand for models with dissipation in the cold medium
proportional to the covering factor $L_{\rm d}/L_{\rm h}=60C$ (see section~\ref{sec:remoteclouds}).   
The assumed distances from the centre are 22.1, 13.6, 9.0, 5.9, 4.0 
times the size of the hot plasma, respectively from left to the
 right hand side.
The Thomson optical depth is fixed to unity, the reflector albedo is $a$=0.1,
the reference inclination angle for reflection (see text) is $i$$=$$30^{\circ}$.
The corresponding covering factors can be estimated from Fig.~\ref{fig:rgdec}. \label{fig:spher_ldproptoC}}
\end{figure}

\begin{figure}
\centerline{\epsfig{file=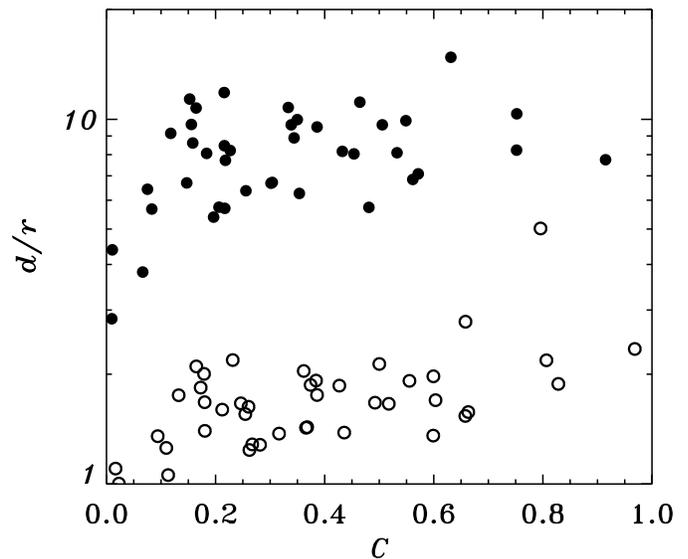,width=8.5cm,height=7cm}}
\vspace*{0.3cm}
\caption{The same $R$-$\Gamma$ data as in Fig.~\ref{fig:spher_ldproptoC}, 
transposed into the $d/r$ vs $C$ plane.
The position of the open circles was determined 
assuming no dissipation in the cold phase ($L_{\rm d}/L_{\rm h}$$=$0),
 while $L_{\rm d}/L_{\rm h}$$=$60$C$ was assumed for the filled circles.
The Thomson optical depth is fixed to unity, the reflector albedo is $a$$=$0.1,
the reference inclination angle for reflection (see text) is $i$$=$$30^{\circ}$.
Data with $R$=0 were excluded (see text).\label{fig:dsrc}}
\end{figure}
\begin{figure}
\centerline{\epsfig{file=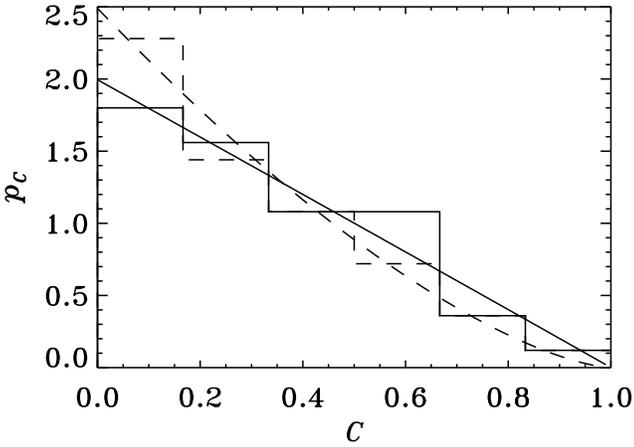,width=8.5cm,height=6cm}}
\vspace*{0.3cm}
\caption{
The observed distribution of the covering factors.
The histograms were determined using the data  (Zdziarski et al. 1999). 
The lines show fits to histograms 
assuming $p_{C}(C)=(1+\alpha)(1-C)^{\alpha}$.
$\alpha$ best fitting values are  $1$ and $1.5$ respectively in the passive case (solid lines, $L_{\rm d}/L_{\rm h}$=0) and in the active clouds case (dashed lines, $L_{\rm d}/L_{\rm h}=60C$).   
The Thomson optical depth is fixed to unity, the reflector albedo is $a$=0.1,
the reference inclination angle for reflection (see text) is $i$=$30^{\circ}$.
\label{fig:pdec}}
\end{figure}

\section{Clouds with internal dissipation}\label{sec:remoteclouds}


\begin{figure*}
\centerline{\epsfig{file=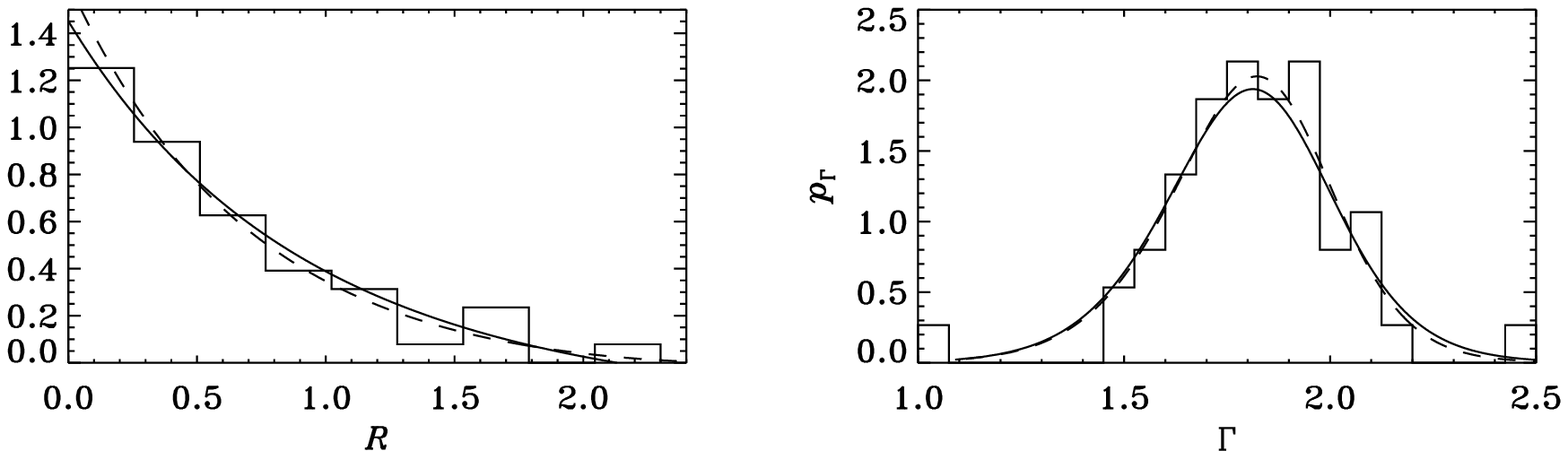,width=17.5cm,height=5cm}}
\caption{Comparison of the predicted spectral parameters distribution with
the observed one.
Left panel: the $R$ distribution, right panel the $\Gamma$ distribution.
The histograms show the observed distributions in the sample of
 Zdziarski et al. (1999). Solid lines shows the model predictions for $L_{\rm d}$=0, $d$=1.7 $r$.
The dashed line show the model predictions for $L_{\rm d}/L_{\rm h}=60 C$ and $d$=8.5$r$ (see section~\ref{sec:remoteclouds}). The assumed distributions of covering factors 
are those shown in Fig.~\ref{fig:pdec}.
 The Thomson optical depth is fixed to unity, the reflector albedo is $a$=0.1,
the reference inclination angle for reflection (see text) is $i$=$30^{\circ}$.
\label{fig:distrib}}
\end{figure*}

\subsection{Dissipation process}

There may be some mechanism providing a substantial
 dissipation in the cold medium, even if we consider spherical accretion. 
An example of such mechanism is turbulent dissipation at
 shocks (Chang \& Ostriker 1985) that are likely to
 form due to the development of spatial
 instabilities (Kovalenko \& Eremin 1998).
It is out of the scope of this paper to discuss or model 
accurately the dissipation mechanism.
The only assumption that we will make about the dissipation process,
is that the total relative power dissipated in the cold phase 
grows linearly with the covering factor i.e.: 
\begin{equation}
\frac{L_{\rm d}}{L_{\rm h}}=\eta C \label{eq:bigprescription}.
\end{equation}
 
It is indeed quite natural to assume that dissipation increases with the
 amount of matter in the cold phase. This is also required by the existence 
of the $R$-$\Gamma$ correlation. Our prescription consists in choosing 
the simplest (linear) function enabling to produce a $R$-$\Gamma$ correlation.
The $\eta$ coefficient  represents the highest relative dissipation
 possible in the cold phase. It is obtained for a fully covered source.
In the following, we will assume that $\eta$ is a (nearly) universal
 constant. We will consider it as a free parameter and  try to get some
constraints with the data.

\subsection{Effects of internal cloud dissipation on $\Gamma$ and $R$} \label{sec:gamadis}

It is obvious that increasing the internal 
dissipation parameter $\eta$, 
will have a similar effect on $\Gamma$, as decreasing the distance ratio $d/r$.
In both cases, the cooling soft radiation flux entering the hot phase becomes
 larger, the amplification factor diminishes and the spectrum softens.

As the effects of both parameters almost compensate, 
very similar $\Gamma (C)$ curves may be obtained for very 
different values $\eta$ and $d/r$ provided they are adjusted
 together in order to keep $\Gamma (C)$ unchanged.
One can derive a simple relation between $\eta$ and $d/r$ enabling
 to conserve $\Gamma (C)$.
 Indeed, expression~\ref{eq:spherampl} for the amplification factor
 can be rewritten using the prescription of 
equation~\ref{eq:bigprescription}:

\begin{equation}
A=\frac{1- \left[ 1-\xi \left( 1+\eta/2\right) \right]} {\xi\left(1+\eta/2\right)C}.
\label{eq:amplificationsimp}
\end{equation}

For simplification, in this expression, second order effects due to reflection
have been neglected (i.e. we set $a$$=$0).
Equation~\ref{eq:amplificationsimp} shows that at first order, the function
 $A(C)$ and thus $\Gamma(C)$ will remain unchanged provided that the quantity:
\begin{eqnarray}
\xi_{0}&=&\xi(1+\eta/2)\nonumber \\
       &=&\left(1-\sqrt{1-\left(\frac{r}{d}\right)^2}\right)(1+\eta/2)\label{eq:constantgammac},
\end{eqnarray}
is kept constant.  $2\pi\xi_{0}$ is the value of the solid angle subtended by
the hot phase as seen from the reflector that provide the same 
$\Gamma(C)$ curve without internal dissipation in the cold phase ($\eta$$=$0).

The solid lines, in the right panel of Fig.~\ref{fig:rgdec}, show different 
$\Gamma(C)$ curves obtained for different distance ratios $d/r$ and no cloud 
dissipation ($\eta$$=$0, see section~\ref{sec:gdec}).
For each solid line, we computed a model \emph{with} internal
 dissipation in the clouds shown in dashed line.
 We fixed  $\eta$$=$60 (this value will be shown to fit the $\fuvx$ 
distribution in section~\ref{sec:cas25}),
 and $d/r$ was computed according to equation~\ref{eq:constantgammac},
with $\xi_{0}$ determined from the distance parameter of the corresponding
non dissipative model. Solid and dashed curves are very close from each other. 
There are only minor differences 
due to effects of reflection.
The global properties of the $\Gamma$ vs $C$ relation are thus unchanged
when dissipation in the cold clouds is taken into account. 
Only the distance-scale is different.

 From equation~\ref{eq:reflection}, one can see that $R$ is
 fully independent of $\eta$. 
And, as quoted in section~\ref{sec:closeclouds},
 $R$ is almost independent of $d$, specially
 when $d$ is large. This can be seen on the left panel of
 Fig.~\ref{fig:rgdec}, the dashed curves show the $R$ versus $C$ 
relation for different distances ($d/r$$>$4), corresponding to those 
of the right panel. They are almost indistinguishable from each other.
The model will always produce
 the correct range of $R$ values.

Then, the shape
 of the predicted $R$-$\Gamma$ correlation at constant $d/r$, appears
 to be nearly unchanged provided that $d$ and $\eta$ are tuned 
simultaneously according to equation~\ref{eq:constantgammac}.
In Fig.~\ref{fig:spher_ldproptoC}, the dashed lines show the model
 $R$-$\Gamma$ correlation for $\eta$$=$60.
 One can see that the observed correlation is then reproduced 
for reflector distances in the range $4r<d<14r$. At the lowest limit, the strong dissipation in the cold phase cools the plasma
down to very low temperatures and the spectra are too soft,
 on the other hand, large $d$ values produce too hard spectra.

\subsection{Effects of internal cloud dissipation on the $C$ and $d/r$ distributions}

When dissipation is assumed, the data points,
 considered in the $d/r$-$C$ plane,  shift
 toward larger $d/r$. 
 The shift in distance can be estimated using 
 equation~\ref{eq:constantgammac}.
Indeed, we know from section~\ref{sec:closeclouds} that for $\eta$$=$0
the average distance is $d/r$$=$1.7. 
  Thus, fixing
\begin{equation}
 \xi_0=1-\sqrt{1-(1/1.7)^2}\sim1/5,
\end{equation}
 in equation~\ref{eq:constantgammac}, provides the general relation
 between $\eta$ and $d$ which enables to fit the
 observed $\Gamma$ distribution. 
In the limit of large distances, 
(effectively 
$d>3r$) expression~\ref{eq:constantgammac} with $\xi_0$$=$1/5 becomes
\begin{equation}
\frac{d^{2}}{r^{2}}\sim\frac{5}{4}\eta\label{eq:chir}.
\end{equation}

This can be understood geometrically as being (for large $d$) 
a relation enabling a constant average soft photon input in 
the hot phase when varying $d$ (or $\eta$).
 It may also tell us that the dissipation process has to 
provide a fixed amount of dissipation per unit surface of reflector.
The numerical constant
in front of the relation is close to unity. This 
indicates that we need nearly the same amount of dissipation
 per unit surface, both in the reflector and the hot medium.

The data distribution in the $d/r$-$C$ plane for $\eta$$=$60, is shown in Fig.~\ref{fig:dsrc}.
 At first sight, the points are  simply shifted toward 
larger $d/r$ as compared to the passive case.
 The average distance is $<$$d/r$$>$=8.5, in agreement with equation~\ref{eq:chir}. The root mean square spread around the average is 44 per cent.
 
A more detailed inspection of Fig.~\ref{fig:dsrc} reveals that the data 
points are slightly shifted toward the left (i.e. lower $C$) compared 
to the passive case. This shift is due to the slight dependence of $R$ on $d/r$.
It affects the $C$-distribution shown in Fig.~\ref{fig:pdec}. 
Equation~\ref{eq:pdec1} provides only a poor approximation to
 the observed $C$-distribution with dissipation.
Instead, the following function:
 \begin{equation}
p(C)=(1+\alpha)(1-C)^{\alpha}\label{eq:pdec2},
\end{equation}
with $\alpha$=1.5, gives a good representation of the distribution (see Fig.~\ref{fig:pdec}).

We fitted the observed $p(C)$ distribution, for different values of $\eta$,
 with the function~\ref{eq:pdec2}
 and $\alpha$ as a free parameter.
 The resulting best fitting values are close to unity as long as $\eta$
 is low $<1$, for larger $\eta$, $\alpha$ converge quickly toward 1.5.
 For any $\eta>$10 the $C$-distribution is well
 represented by equation~\ref{eq:pdec2} with $\alpha\sim$1.5.
We checked that these results are not significantly affected
 by statistical errors due to a particular bining of the $C$-distribution. 

In all cases the derived $p(C)$ distribution provided a reasonable agreement 
with the $R$ and $\Gamma$ distributions, as long as $d/r$ 
is set to the average value given by the data. 
Fig.~\ref{fig:distrib} illustrates the case $\eta$=60 with 
$d/r$=8.5, comparing
 the derived $R$ and $\Gamma$ distributions with the data 
in a way similar to that of section~\ref{sec:rgdis}.
 As expected, they are very well reproduced.

\subsection{$\fuvx$ distribution \label{sec:cas25}}

\begin{figure}
\centerline{\epsfig{file=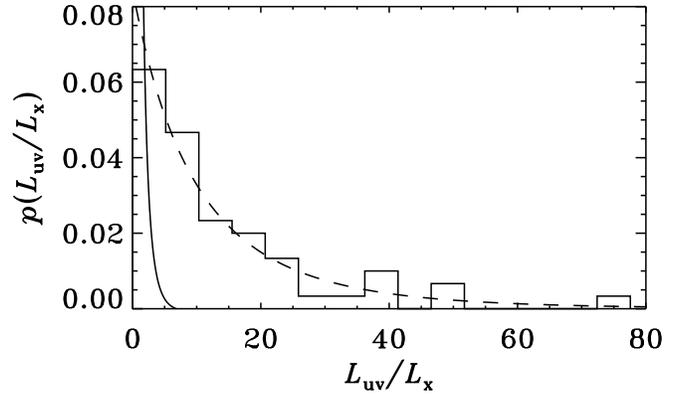,width=8.5cm,height=5cm}}
\caption{Comparison of the predicted $\fuvx$ distribution with
 the observed one. The histogram shows the observed distribution
 in the sample of Walter \& Fink (1993). The curves show the model 
distribution computed according to equation~\ref{eq:luvtolx} and assuming
the distributions of covering factors shown in Fig.~\ref{fig:pdec}.
The solid line is for a reflector distance $d$=1.7$r$
 without any dissipation in the clouds: additional sources 
of UV radiation are required to explain the observed distribution (see section~\ref{sec:fuvx}).
 Dashed line is for $d$=8.5$r$ and a relative dissipation in the clouds such
 that $L_{\rm d}/L_{\rm h}=60C$ (see section~\ref{sec:remoteclouds}): the UV emission can be understood as being totally emitted by the cold clouds.
In both cases optical depth is $\tau$=1, the cloud albedo is $a$=0.1
 (see section~\ref{sec:estimates}).}

\label{fig:luvtolx}
\end{figure}

The observations provide an upper limit on $\eta$.
 Indeed in the limit of large $\eta$, 
this parameter is roughly comparable to the largest
 UV to X-ray luminosity ratio predicted by the model.
As a consequence, $\eta$ can not be much larger than the maximum observed
 UV to X fluxes ratios.
 It turns out (see e.g. Fig.~\ref{fig:luvtolx}) that the 
largest observed UV to X ratios are in the range 10-100.
This observational fact constrains $\eta<100$. As a consequence of
 equation~\ref{eq:chir}, this limit 
 indicates that the reflector cannot be at a distance
 larger than $\sim13r$.

Fig.~\ref{fig:luvtolx} shows that the observed $\fuvx$
distribution in Seyfert~1 galaxies is well represented
 by the model with $\eta$=60 (dashed line). This result
 is not very sensitive to $\eta$ as
 long as it keeps being comparable to the largest observed $\fuvx$ 
values. Reasonable agreement with the observed distribution can be
 achieved for 
$40<\eta<80$. The predicted distribution 
is insensitive to reflector distance as long as $d>3r$.

We stress again that  the distribution of the observed $\fuvx$ 
does not  provide any lower limit on $\eta$, which can be much 
smaller than 60, the limiting case $\eta$=0 was studied in 
section~\ref{sec:closeclouds}. Then, however,
 external sources of soft radiation are required in order to explain
 the important amount of sources with
 large $\fuvx$ ratios.

We conclude that fixing the universal relative dissipation
 to be roughly $L_{\rm d}/L_{\rm h}\sim 60C$, and the  distance $d\sim 8.5r$, 
enable to explain the $R$ and $\Gamma$ distribution, their correlation and,
 independently the observed distribution of $\fuvx$. 
This result is not very sensitive to the exact value of the parameters:
a qualitative agreement can be achieved in a quite large 
region of the parameter space.

\section{Additional considerations}\label{sec:consideration}

\subsection{Absorption features}\label{sec:absfea}

The observed column density of absorbing cold material
 on the line of sight of Seyfert 1 galaxies
 is usually quite weak ($N_{H}\simlt10^{21}$ cm$^{-2}$).
 As already quoted by Celotti et al. (1992) this may indicate that
 the cloud column density is low. A larger amount of absorbing material would 
induce a characteristic  hardening of the spectral slope in the soft X-ray,
 since the lower energy photons are more likely to be absorbed
than harder ones.
 Such an absorption feature would be easily detected.  

However, our comparisons with the data have shown that,
 at least for some sources, large covering factors are required.
A low absorbing column density is not compatible with a large fraction of 
primary radiation being intercepted (up to 80--90~per cent).
  A possibility would be, on the contrary, to have  the clouds  optically \emph{very}
 thick (i.e. highly absorbing, $N_{H}\sim 10^{25}\quad \mathrm{cm}^{-2}$).
 So that, the fraction of escaping X-ray luminosity 
intercepting the clouds would be fully blocked, while the rest would escape
without any spectral alteration.

The constraints on the clouds column density 
are less stringent at low covering factor.
 Then, the effects of absorption on the observed spectrum 
are weak in any case.
The observed $R$ and $\Gamma$ distributions suggest that    
 most of the sources are observed with a low covering factor.
 According to the distribution~\ref{eq:pdec1},
 75~per cent of the observed objects would have $C<0.5$. 
Thus, for most of the observed sources the cloud optical 
thickness may be low.
When the amount of darkening material is increased, it is likely 
that the additional matter spreads in the three dimensions, so that
 its radial optical depth grows together with the surface covered.

Still, the absorption effects
 might be observable in some sources,
 with large covering fractions i.e. very
steep power law spectrum.
At low energy ($<$ 10 keV) the radiation escapes only 
through the uncovered parts of the source.
Above 20 keV the clouds are transparent to radiation,
 the observed 20-30 keV flux could be larger
than the extrapolation of the 2-10 keV one, by nearly an order of magnitude. 
This would induce a strong hardening of the spectrum similar
 to what is observed in some obscured Seyfert 2 galaxies.
Since such a characteristic is specific 
to the quasi-spherical accretion scenario, a detailed  
investigation of the soft gamma-ray spectrum
 of the softest Seyfert 1 galaxies would be
of great interest.
However, even at high energy,
 the radiation still suffers Compton down 
scattering on the cold matter.
 If the cloud Thomson optical depth is large, 
the energy loss is important and the down scattered 
radiation is absorbed before escaping from the cloud.
This may limit the observability of the absorption features.

\subsection{Is the observed $C$-distribution due to selection effects ?}\label{sec:obscuration}

If, as suggested in section~\ref{sec:absfea}, the clouds are
highly absorbing, it may arise that in some objects, the X-ray 
source is completely hidden, either because one large cloud
 is on the line of sight, or because 
 a particular cloud distribution happens to fully 
cover the X-ray source as seen from earth.   
Obviously such sources would be very unlikely to be detected.
Such an effect may explain that according to the $C$ 
distribution derived  from the data, 
we observe much more sources with a low covering factor
 than highly covered ones.

Indeed, for a nearly homogeneous intrinsic $C$-distribution,
and if, for the observer, the angular size of an individual cloud, 
is much larger than the hot source size
 (i.e. in the limit of extended clouds at large distances), the 
probability of detecting a source decreases exactly 
like its uncovered fraction  $(1-C)$.
In the general case, the exact $C$-distribution depends on the average
 number of clouds, and their distance and size distributions.
 In addition, both size and number are likely to be a function of $C$.
 Depending on the details, one can reasonably get
a distribution decreasing more or less
according to $(1-C)^{\alpha}$ with $\alpha$ in the required range.

This kind of extinction effects are more likely to occur if
 the cloud distance is large.
Simple geometrical considerations show that a complete extinction is
possible only when $(r/d)^{2}>(2c-c^{2})^{-1/2}$.
 Even above this limit,  
strong departure from spherical geometry 
may be required. 
Such effects are thus unlikely to be important at low $d/r$ 
(i.e.if there is no dissipation in the clouds).
 On the other hand, the observed $C$-distribution may be
strongly affected by such selection effects if the clouds are at a
 larger distance, i.e. $d/r>3$,
 as considered in section~\ref{sec:remoteclouds}.

In the case where the source of primary emission is hidden,
it might be detected in the X-rays through 
the reflected radiation coming from the clouds themselves. 
Thus, if obscuring effects are
 important, we can expect to detect some very faint sources with
 infinite reflection coefficient. Such detections have not been reported.
However, at least one previously known and usually
 bright source (NGC4051; Guainazzi et al. 1998) has 
  been observed in a very faint state dominated by reflection.
 Such spectrum can be interpreted as being due to a obscuring event.

The obscuring clouds of the present 
model cannot be identified with the obscuring material 
that hide the broad line region (BLR) in Seyfert 2
 galaxies. The BLR is
 at large distance from the black hole ($>10^3$ gravitational radius $R_{g}$),
 so the obscuring material is, at least, at a similar distance. 
On the other hand our clouds are constrained to be 
in the very inner part of the accretion flow ($<100$ $R_{\mathrm g}$).

\begin{figure}
\centerline{\epsfig{file=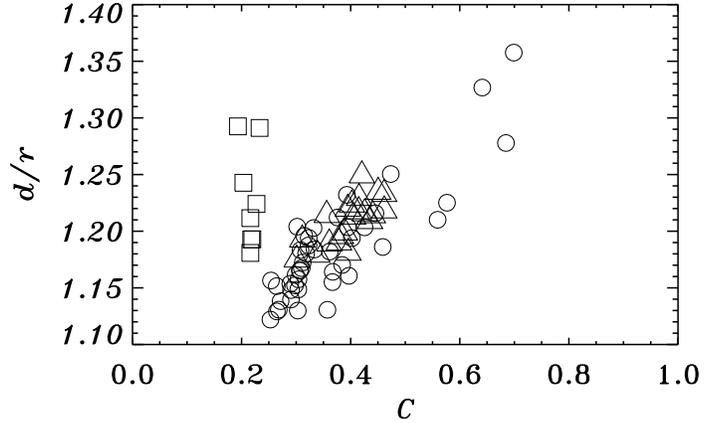,width=8.5cm,height=5cm}}
\vspace*{0.3cm}
\caption{The  $R$-$\Gamma$ data of Gilfanov et al. (2000) for galactic black holes, transposed into the $d/r$ vs $C$ plane. Circles, triangles and squares
 represent data for Cygnus X-1, GX 339-4 and GS 1354-644 respectively.
There is no dissipation in the cold phase ($L_{\rm d}/L_{\rm h}$=0).
The Thomson optical depth is fixed to unity. The reflector albedo is larger for GBH than AGN (see MBP), we set $a$=0.15.
Following GCR00, the reference inclination angle for reflection (see text) is set to $i$=$50^{\circ}$.
The coefficients of B99b ($\gamma$=2.33, $\delta$=1/6) relevant for GBH are used 
in equation~\ref{eq:total}.
\label{fig:dsrgbh}}
\end{figure}

\subsection{Galactic black holes}

Since GBH sources present hard X-ray spectral
 properties which are very similar to those of Seyfert 1
 galaxies, this model 
can, in principle, be applied to such objects.
We compared the predicted correlation with the data given by GCR00.
In these comparisons, the $\Gamma$ versus $A$ relation was approximated by
 equation~\ref{eq:total} with coefficient $\gamma$ and $\delta$ appropriate for GBH. We used both
 coefficients given by B99b ($\gamma$=2.33, $\delta$=1/6) and
 MBP ($\gamma$=2.19, $\delta$=2/15).
In both cases, we found a reasonable agreement 
 with the observed $R$-$\Gamma$ correlation 
for $d$ in the range 1.1--1.4, without dissipation in the clouds.
Fig.~\ref{fig:dsrgbh} shows the data of GCR00
 represented in the $d/r$-$C$ plane. The average $d/r$ value is 1.2. The rms $d/r$  spread is about 4 per cent, much smaller than in the extra-galactic case.

The distribution of spectral parameters is found to be inconsistent with a
covering factor distribution of the type given by~\ref{eq:pdec2}.
 The sample of data of 
GCR00 contains only 3 sources
 (Cygnus X-1, GX 339-4 and GS 1354-644) with
 numerous observations of each source. 
This situation differs from the Seyfert sample where
we have a larger amount of sources (23) with less observations
 of each sources (2 on average).
In GBH, we trace the physical evolution 
of a few particular sources, while in Seyfert galaxies, we trace the properties
 of the whole class.
In Seyfert galaxies also, selection effects are more
 likely to bias the distribution.
Therefore, the source statistic is expected to be different.

It is not clear if such a spherical model could 
explain the complex temporal behavior of galactic sources,
 whose interpretation generally requires the presence of a disk 
very close to the central object. Actually, the striking spectral 
similarities between GBH sources and Seyfert galaxies
(similar hard X-ray spectra, correlation R-$\Gamma$), are probably due
to the universal properties of Compton self-regulated plasma.
The underlying heating process and geometry could be however
very different.

 \subsection{The iron line profile}

We focussed on the properties of the continuum spectra
 and did not attempt to model any line emission. 
It is however worth mentioning  
the iron K$\alpha$ fluorescence line at 6.4 keV. 
Such lines are indeed commonly observed in Seyfert galaxies 
(Nandra \& Pounds 1994; Nandra et al. 1997).
 They generally present a broad, possibly red-shifted 
profile. 
 The usual interpretation is that the line forms trough illumination of
 the innermost part of an accretion  disc. The line profile then
appears broadened and red shifted due to special and
 general relativistic effects (Fabian et al. 1989; Tanaka et al. 1995). 
It seems however that some alternative mechanisms may 
produce the broad line profiles without a disc (see Karas et al.~2000, and references therein).
Such mechanisms include:
\begin{enumerate}
\item Radiative transfer effects, like
reflection by a ionized medium with an intermediate temperature ($kT\sim 1$ keV). This warm medium could be the transition region located 
between the hot phase and the cold clouds. 
\item  Doppler effects due to cloud motions. 
For example, if the clouds form
 an out-flowing wind. The line is formed by reflection on 
the inner side of the clouds traveling away from the observer.
It is then Doppler red-shifted and broadened due to the spread
in the clouds radial velocities. 
\item Gravitational red-shift, if the size the system is small enough.
\end{enumerate}
\noindent Or a combination of these different effects.
Note also that a recent reanalysis of the {\it ASCA} data (Lubi\`nski \& Zdziarski~2000)
 indicates that in most of sources, the broadening is much 
less important than previously thought.

\subsection{Ionization}

We assumed that the clouds are dense enough so that
the ionization parameter is low ($<10^{3}$) and the ionization effects
 are weak (e.g. \.Zycki et al. 1994).
Then, the cloud albedo is low $a\sim0.1$ and most of the hard X-rays
 impinging on the clouds are absorbed.

For large ionization parameter a ionized skin forms on the irradiated surface of the clouds.
Such ionization may have important effects on 
the radiative equilibrium (Ross, Fabian \& Young 1999; Nayakshin, Kazanas \& Kalman 2000) and change many observational parameters
such as the hard X-ray spectral slope,
 the amplitude of the reflection component or the UV to X-ray
 luminosity ratio. 
 Despite we restricted
 our considerations to neutral clouds, most of our analytical
 estimates, given in section~\ref{sec:estimates}, may be used to study 
ionized cases, simply by setting the cloud albedo to a convenient value.

Only equation~\ref{eq:spherR}, giving the reflection amplitude,
would breakdown. Indeed, the $R$ parameter, as measured in spectral fits
 gives the amount of reflection on a nearly neutral matter.
Reflection on strongly ionized material produces a spectrum which is 
indistinguishable
from the primary emission. 
Our estimate gives the truly reflected luminosity both on neutral
 and ionized matter.
Equation~\ref{eq:spherR} thus overestimates the effective $R$ when
 ionization is important.

\section{Conclusion}

We assumed a
 spherical distribution of reflecting/absorbing matter surrounding 
the central source of primary radiation in Seyfert galaxies. 
We showed that such a situation is consistent with the $R$-$\Gamma$ data.
The model successfully reproduces the range of observed spectral
 slopes and reflection amplitudes, the $R$-$\Gamma$ correlation and 
the individual $R$ and $\Gamma$ distributions. 
The same model is also consistent with the $R$-$\Gamma$ data
 of galactic black holes.

The two main quantities controlling the value of the spectral parameters are
the average cloud distance to the hot plasma size ratio $d/r$,
 and the covering factor $C$. 
The observations of Seyfert galaxies put the following constraints on the model parameters:
\begin{enumerate}
\item The data indicate that the distance $d/r$ of the cold material 
 is comparable in all objects.
On the other hand, wide changes in the 
covering factor $C$ from source to source 
would be responsible for the observed spectral differences. 
 \item The observed distributions of spectral parameters
 require the distribution of covering factors to decrease with $C$.
 A $C$-distribution of the form
 $p(C)=(1+\alpha)(1-C)^{\alpha}$ with $\alpha$ in the range 1--1.5,
 gives a good description of the data.
If $d$ is large ($d>3r$) this kind of distribution
 may be explained by darkening effects.
Otherwise, if $d<3r$, it is more likely that the observed 
$C$-distribution reflects the intrinsic one.  
\item Without dissipation in the cold phase:
the clouds are constrained to be in the immediate 
vicinity of the central hot plasma.
The data indicate an average relative distance $<$$d/r$$>$=1.7
with a small spread in distance from source to source.
The index $\alpha$ of the $C$-distribution is 1.
\item If there is some internal dissipation in the soft phase,
 growing proportionally to $C$ ($L_{\rm d}/L_{\rm h}=\eta C$):
 the cloud distance may be larger,
 depending on the amount of dissipation in the cold phase.
For $d>3r$, the average distance required by the data 
grows with $\eta$, like $d/r\sim\sqrt{1.25 \eta}$.
\item  In any case 
the distance is constrained to
 be lower than $d\sim13r$, by the maximum observed 
UV to X flux ratios.
 Fixing $d\sim8.5r$ (and $\eta\sim 60$) enables us 
to understand
the bulk of the UV emission as being emitted by the spherical 
reflector itself. Then, the index of the $C$-distribution,
 derived from the X-ray data,  is $\alpha$$=$1.5.
The distribution of $\fuvx$ ratios derived from this $C$-distribution 
is consistent with the observed one.
If the cloud distance is lower than 8.5$r$, the $\fuvx$ data suggest that 
part of the observed UV flux is emitted by some additional 
source, external to the cloud system.

\end{enumerate}

This simple  model, with a limited number of 
parameters, enables us to understand
 the bulk of the phenomenological
 properties of the continuum emitted in Seyfert 1 galaxies.
 We believe this is a strong support in favor of a spherical inner 
accretion flow in these sources.

\section*{Acknowledgments}

This work was supported by a grant 
from the Italian MURST (COFIN98-02-15-41).
 I am grateful to Suzy~Collin, Elisabeth~Jourdain and Laura~Maraschi
 for a careful reading of the manuscript and many useful suggestions.
 I am indebted to Andrei~Beloborodov for critical comments
and to Pierre-Olivier~Petrucci for checking
the calculations and many discussions
 on the observation of Seyfert galaxies.
 I also thank an `anonymous' referee 
for several important suggestions, as well as 
Andrej~Zdziarski and Marat~Gilfanov
for providing me with the $R$-$\Gamma$ data.


\begin{thebibliography}{}

\bibitem[]{ac00}
Abrassart A., Czerny B., 2000, A\&A, 356, 475 (AC00)

\bibitem[ ]{bel2}
Beloborodov A. M., 1999a, ApJ, 510, L123

\bibitem[ ]{bel4}
Beloborodov A. M., 1999b, in Poutanen J., Svensson R., eds,
ASP Conf. Series Vol. 161, High Energy Processes in Accreting Black Holes.
Astron. Soc. Pac., San Francisco, p. 295 (B99b)

\bibitem[]{bel00}
Beloborodov A.M., 2001, ``Accretion disk models for luminous black holes'',
 33rd COSPAR Assembly, to appear in Adv. Space Research, astro-ph/0103320 

\bibitem[ ]{bk77}
Bisnovatyi-Kogan G.S., Blinnilov  S.I., 1977, A\&A, 59, 111

\bibitem[ ]{cfr92}
Celotti A., Fabian A.C., Rees M.J., 1992, MNRAS, 255, 419

\bibitem[ ]{co85}
Chang K.M., Ostriker J.P., 1985, ApJ, 288, 428

\bibitem[]{ccdz96}
Collin-Souffrin~S., Czerny~B., Dumont~A.-M., Zycki~P. T., 1996, A\&A, 314, 393

\bibitem[]{cacdm00}
Collin~S., Abrassart~A., Dumont~D., Mouchet~M., in proc. "AGN in their Cosmic Environment", Eds. B.~Rocca-Volmerange \& H.~Sol, EDPS Conf. Series in Astron. \& Astrophysics, in press, astro-ph/0003108

\bibitem[ ]{c92}
Coppi P.S., 1992, MNRAS, 258, 657

\bibitem[]{cd98}
Czerny B., Dumont~A.M., 1998, A\&A, 338, 386

\bibitem[ ]{frsw89}
Fabian A.C., Rees M.J., Stella L., White N.E., 1989, MNRAS, 238, 729 

\bibitem[ ]{gf91}
George I.M., Fabian A.C., 1991,  MNRAS, 249, 352


\bibitem[ ]{ghm94}
Ghisellini G., Haardt F., Matt G., 1994, MNRAS, 267, 743


\bibitem[ ]{gilf2}
Gilfanov M., Churazov E., Revnivtsev M., 2000, 
in Proc. 5th CAS/MPG Workshop on High Energy Astrophysics, in press, astro-ph/0002415 (GCR00)  


\bibitem[ ]{Guain98}
Guainazzi M. et al., 1998, MNRAS, 301, L1

\bibitem[ ]{gr88}
Guilbert P.W., Rees M.J., 1988, MNRAS, 233, 475

\bibitem[ ]{hm93}
Haardt F., Maraschi L., 1993, ApJ, 413, 507


\bibitem[ ]{hmg94}
Haardt F., Maraschi L., Ghisellini G., 1994, ApJ, 432, L95

\bibitem[ ]{kcaa00}
Karas V., Czerny B., Abrassart A., Abramowicz M.A., 2000, MNRAS, 318, 547

\bibitem[ ]{ke98}
Kovalenko I.G., Eremin M.A., 1998, MNRAS, 298, 861

\bibitem[ ]{kmz94}
Krolik J.H., Madau P., \.Zycki P.T., 1994, ApJ, 420, L57

\bibitem[ ]{kbr96}
Kuncic Z., Blackman E.G., Rees M.j., 1996, MNRAS, 283, 1322

\bibitem[ ]{kcr97}
Kuncic Z., Celotti A., Rees M.J., 1997, MNRAS, 284,717

\bibitem[ ]{lz00}
Lubi\'nski P., Zdziarski A.A., MNRAS, submitted, astro-ph/0009017

\bibitem[ ]{mz95}
Magdziarz P., Zdziarski A. A., 1995, MNRAS, 273, 837

\bibitem[ ]{mj00}
Malzac J., Jourdain E., 2000, A\&A, 359, 843

\bibitem[ ]{mbp00}
Malzac~J., Beloborodov~A., Poutanen~J., MNRAS in press, astro-ph/0102490

\bibitem[ ]{np94}
Nandra K., Pounds K.A., 1994, MNRAS, 268, 405

\bibitem[ ]{ngmty97}
Nandra K., George I.M., Mushotzky R.F., Turner T.J., Yaqoob T., 1997, ApJ, 477, 602

\bibitem[ ]{nkk}
Nayakshin S., Kazanas D., Kallman T.T., 2000, ApJ, 537, 833

\bibitem[ ]{pet00}
Petrucci P.O. et al., 2000, ApJ, 540, 131


\bibitem[ ]{pou98}
Poutanen J., 1998, in Abramowicz M. A.,  Bj\"ornsson G., Pringle J., eds,
Theory of Black Hole Accretion Disks. Cambridge Univ. Press, Cambridge, p. 100

\bibitem[ ]{pkr97}
Poutanen J., Krolik J. H., Ryde F., 1997, MNRAS, 292, L21

\bibitem[]{rfy99}
Ross R.R., Fabian A.C., Young A.J., 1999, MNRAS, 306, 461

\bibitem[ ]{mcmethod}
Stern B., Begelman M.C., Sikora M., Svensson R., 1995, MNRAS, 272, 291

\bibitem[ ]{sle76}
Shapiro S.L., Lightman A.P., Eardley D.M., 1976, ApJ, 204, 187

\bibitem[ ]{sd96}
Smith D.A., Done C., 1996, MNRAS, 280, 355

\bibitem[ ]{st80}
Sunyaev R.A., Titarchuk L.G., 1980, A\&A, 86, 121

\bibitem[ ]{t95}
Tanaka Y. et al., 1995, Nature, 375, 659


\bibitem[]{WF93}
Walter~R., Fink~H.~H., 1993, A\&A, 274, 105 

\bibitem[ ]{WZ00}
Wardzinski G., Zdziarski A. A., 2000, MNRAS, 314, 183

\bibitem[ ]{zdz97}
Zdziarski A. A., Johnson W. N., Poutanen J., Magdziarz P., Gierli\'nski M.,
1997, in Winkler C., Courvoisier T. J.-L., Durouchoux Ph., eds,
Proc. 2nd INTEGRAL Workshop, The Transparent Universe, ESA SP-382.
  ESA, Noordwijk.  p. 373

\bibitem[ ]{zls97}
Zdziarski A. A., Lubi\'nski P., Smith D. A., 1999, MNRAS, 303, L11 (ZLS99)

\bibitem[ ]{zyc94}
\.Zycki P. T., Krolik J. H., Zdziarski A. A., Kallman T. R., 1994, ApJ, 437, 597
\end{thebibliography}
\end{document}